# Modulation of Spin Angular Momentum of Emission in Symmetric 1D Plasmonic Crystals by Cathodoluminescence

*Yuxin Yang[1], Izzah Machfuudzoh[1], Qiwen Tan[1], Takumi Sannomiya[1,*]*

[1] Department of Materials Science and Engineering, School of Materials and Chemical Technology, Institute of Science Tokyo, 4259 Nagatsuta, Midori-ku, Yokohama 226-8503 Japan

**Corresponding author**

*Takumi Sannomiya (Email: sannomiya@mct.isct.ac.jp)

**ABSTRACT**

The spin angular momentum (SAM) of light has become a cornerstone of numerous photonic applications, including optical communication and chiral photonics. Because SAM is inherently associated with circularly polarized light (CPL), the ability to modulate CPL in a controlled and efficient manner is essential not only for advancing fundamental studies of light–matter interactions but also for enabling next-generation photonic technologies. However, such modulation is commonly realized by structurally chiral systems, which inherently limits the feasibility of dynamic tuning. Here, we demonstrate that one-dimensional plasmonic crystals (1D PlCs), despite their structural symmetry, can serve as a platform for controllable CPL generation. By employing an electron beam in scanning transmission electron microscopy (STEM), we coherently excite transition radiation and emission from 1D PlC modes. Their interference produces energy- and momentum- (emission angle-) resolved CPL, which clearly reveals its dispersion and spatial dependence at the nanoscale, providing direct guidance for its manipulation and offering insights into the design of plasmonic devices including the phase information. Furthermore, interference with surface plasmon polariton scattering at the structural boundary enables the efficiency modulation of CPL generation via the excitation position along the terrace.

# Introduction

The spin angular momentum (SAM) of light constitutes a fundamental and versatile degree of freedom for encoding and manipulating optical information in modern photonics. It takes two discrete values, $+\hbar$ and $-\hbar$, where $\hbar$ is the reduced Planck constant, corresponding to left- and right-handed circular polarization (LCP and RCP), respectively. This intrinsic binary nature endows SAM with distinct advantages as an information carrier, notably robustness against environmental perturbations[1,2] and straightforward compatibility with existing optical systems.[3–5] Consequently, SAM has become a cornerstone of numerous photonic applications, such as optical communication,[6,7] quantum information processing,[8,9] and chiral photonics[10,11]. Given its fundamental significance and broad applicability, the precise generation and manipulation of the SAM of light are highly crucial. Because SAM is inherently associated with CPL, efficient and flexible CPL modulation is therefore essential not only for advancing fundamental studies of light–matter interaction[12,13], but also for enabling next-generation photonic technologies[14].

Conventional strategies for manipulating CPL typically rely on nanostructures with intrinsic geometric asymmetry, such as asymmetrical metasurfaces with additional angular momentum[15,16], or twisted multilayer stacks to support collective geometrical chirality.[17,18] These structures interact differently with left- and right-circularly polarized excitations, enabling selective enhancement or suppression of one handedness over the other. However, such intrinsic chiral structures come with a major limitation: once fabricated, their chiroptical responses are fixed and lack post-fabrication tunability. This static nature hinders dynamic CPL control, while it is crucial for reconfigurable or spatially programmable photonic devices. In addition, their fabrication often involves complex lithographic processes, limiting the scalability.

To address these limitations, adoption of symmetrical structures to control CPL has emerged as an alternative approach, where symmetry is broken by external factors, including localization excitation and detection, so-called extrinsic chirality. Typical examples include illuminating only a portion of an otherwise symmetric structure with a tightly focused beam,[19–21] selectively probing a local region using a near-field tip,[22,23] and collecting emitted or scattered signals from a limited spatial domain instead of the entire sample[24]. Under these conditions, the intrinsic symmetry of the structure is preserved, but the external excitation or detection scheme imposes spatial asymmetry, thereby activating responses that would be absent or substantially weaker in a fully symmetric configuration. Compared to fixed chiral geometries, symmetrical structures exhibit the advantage of equal responsiveness to both CPL polarizations, thereby enabling controllable and reversible modulation of chiral optical properties depending on the excitation position, photon energy, and detection angle, offering a more flexible route for dynamically manipulating CPL polarization (the SAM of light).[25,26] While most previous studies utilize antenna structures,[22,25] here we focus on one-dimensional plasmonic crystals (1D PlCs), as a simple symmetrical structure involving excitation modes caused by propagating surface plasmon polaritons (SPPs) on periodic metasurfaces, which are capable of generating complex and tunable polarization states.[27–29] To break its symmetry extrinsically, electron beam offers a unique and precise approach owing to its highly localized excitation capability. It can excite sample at the nanoscale beyond the diffraction limit of light[30–33] and deliberately break structural symmetries at targeted positions.[25,34,35] This spatial selectivity is particularly advantageous for investigating position-dependent optical responses in symmetric plasmonic systems, considering the requirement for precise excitation within the size of 1D PlCs.

In this work, we employ scanning transmission electron microscopy (STEM) combined with cathodoluminescence (CL) detection systems to utilize electron beam as localized excitation source and to analyze the chiral optical responses of 1D PlCs. As schematically illustrated in Figure 1a, when electron beam irradiates the metal surface of 1D PlCs, both transition radiation (TR) and emission from SPP-induced 1D PlC mode are generated. By leveraging their interference, we achieve controllable CPL generation, which is observed in both large-terrace (with a terrace width of 420 nm) and small-terrace (with a terrace width of 120 nm) 1D PlC, as shown in the left and right panels of Figure 1b, respectively. By taking advantage of the CPL handedness, we can also analyze the phase of 1D PlC mode with reference to TR. Near the structural boundary, the interference between CPL and SPP scattering induces distance-dependent CPL modulation, thereby providing an additional degree of freedom. All optical signals are collected using a four-dimensional (4D) STEM-CL technique capable of simultaneously acquiring emission angle- and energy-resolved information in all detection angles and photon energies (Figure 1c).[25,35] Details can be seen in Method section.

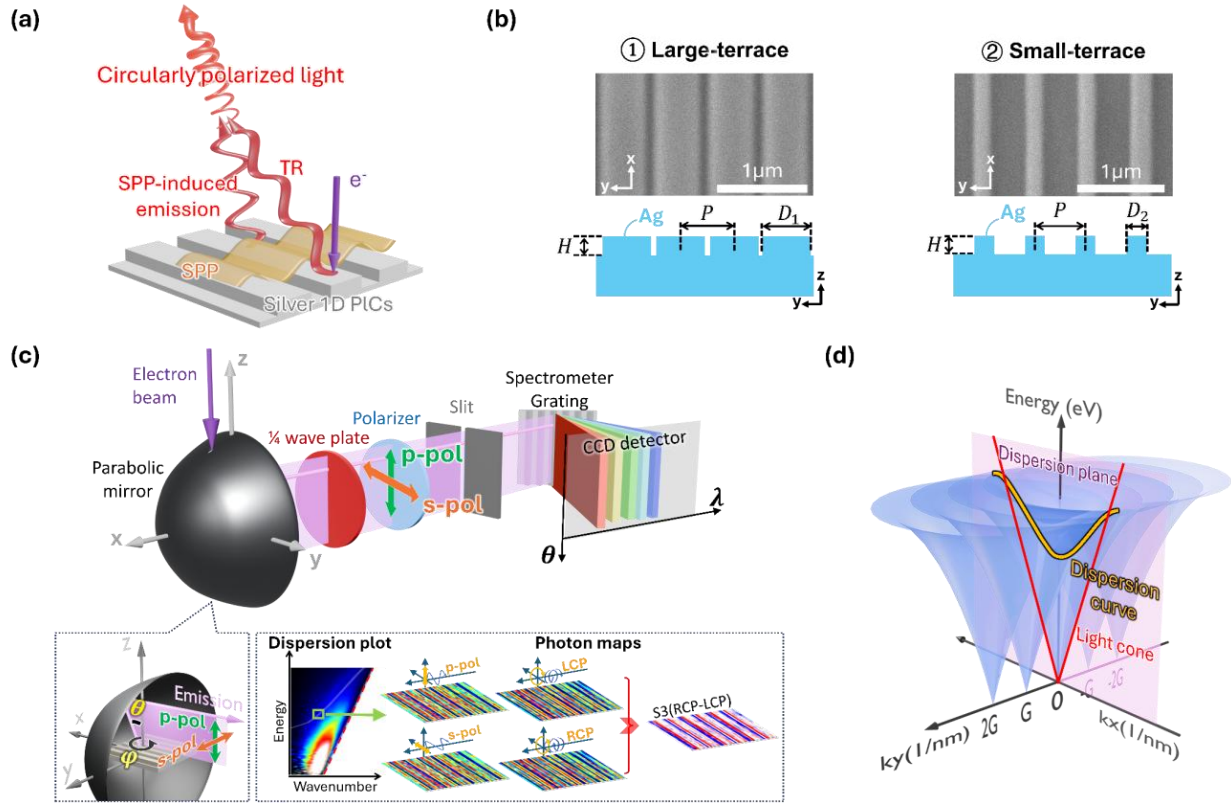

**Figure 1.** (a) Illustration of CPL generation through interference of TR and emission induced by SPP excited by an electron beam. (b) Backscattered electron images of 1D PlCs with large terrace and small terrace samples from a top *xy* view. The geometric dimensions of terrace height ($H = 100$ nm) and periodicity ($P = 600$ nm) are identical for both samples. The large-terrace sample has the terrace width ($D_1$) of 420 nm, and the small-terrace sample width ($D_2$) of 120 nm. (c) Schematics of the experimental setup used to acquire energy- and angle-resolved (wavenumber- or momentum-resolved) CL maps and dispersion relations via an angle-resolved spectrum acquisition system. In the angle-resolved measurements, the polar emission angle range of $0° \leq \theta \leq 90°$ is considered, and the azimuthal angle $\varphi = 0°$ is selected by the mask slit. (d) Schematic diagram of the dispersion relation of light and surface plasmon polaritons (SPPs). The pink plane represents the dispersion plane cross-section where the measured data lies. The blue cones

represent SPP dispersion, while the red lines denote the light line. The projected dispersion curve of the 1D PlC is represented by the yellow line.

## RESULTS AND DISCUSSION

**CPL generation in the 1D PlC with large terrace (420 nm)**

The 1D PlCs are oriented such that the periodicity vector of the sample is perpendicular to the *x*-axis of the collection system (see the bottom left of Figure 1c). This configuration enables the collection of periodic dispersion features from *s*-polarized light[29], with the electric field parallel to the *y*-axis, as illustrated in Figure 1d. The measurable dispersion curve is the $k_y = 0$ cross-section of the dispersion cones shifted by reciprocal lattice vector G, restricted to the region inside the light cone (Figure 1d). This is because the slit in the detection system (Figure 1c) restricts the collected CL to the plane at azimuthal angle $\varphi = 0°$, corresponding to the $k_y = 0$ plane. In addition, only the modes inside the light cone can be detected by the STEM-CL system, because it measures radiative far-field emission, whereas modes outside the light cone are non-radiative and therefore cannot be observed in the present experiment. Figure 2 presents the CL measurement results for the large-terrace sample (Figure 1b, left). Figure 2a and 2b show respectively the dispersion relations of *p*- and *s*- polarized light with the CL intensity integrated over the entire scanning area provides an overview of this 1D PlCs. Here, *p*-polarized light denotes the polarization with the electric field component parallel to the *z*-axis, as illustrated in Figure 1d. In Figure 2a, the strong emission intensity distributes along the light line, indicating that the *p*-polarized emission is mainly caused by TR, which is well known to be polarized normal to the surface plane, corresponding to *p*-polarization.[27,29] Therefore, in Figure 2b, TR can be excluded by using *s*-polarization and only emission from 1D PlC remains in the dispersion plots. The detected emissions in Figure 2b align well with the calculated SPP dispersion curves (solid purple lines), confirming the dispersive features of the SPP-induced 1D PlC mode. Notably, a splitting of the bright pattern along the dispersion curve is also observed, indicating two different 1D PlC modes.

To identify these two 1D PlC modes, an energy line profile (Figure 2c) with a fixed detection angle is utilized to show the field distributions on the sample at different energies, where two modes with distinct charge distributions are observed. At higher energy (indicated by the green dashed line), charges with opposite signs appear at the centers of the terraces and grooves, corresponding to the symmetric (S) mode, whereas at lower energy (orange dashed line), charges are observed at the edges of the terraces, which is the antisymmetric (A) mode.[29] Although S mode should not be detected under the ideal detection condition at azimuthal angle $\varphi = 0°$, the finite slit width used to collect the sufficient signal allows some S mode component to be captured. Because these two modes exhibit different charge distributions relative to the sample structure that result in different mode energies, a bandgap is formed at the $\Gamma$ point ($k_x = 0$ nm$^{-1}$), corresponding to emissions at polar angle $\theta = 0°$.[29] We note that a detection angle of $\theta = 20°$ is used for the line profile of Figures 2c to collect emissions from all excitation modes as efficiently as possible, while remaining sufficiently close to the $\Gamma$ point.

To elucidate how these excitation modes vary with energy and emission angle, the angle-resolved energy line profiles of *s*-polarized light obtained from a single terrace is studied, with 10° intervals (Figure 2d). The solid line represents the theoretically calculated dispersion curve in angular space, converted from the corresponding curve in momentum ($k$)-space (Figure 2b). As shown in Figure 2d, both A mode (observed at terrace edges below the solid pink line) and S mode (observed at terrace center above the solid pink line) shift consistently with calculated dispersion, showing their dispersive feature. This further confirms that both modes originate from SPPs modulated by the periodic structure. A comparison with Figure 2b indicates that the intensities below the dispersion line in Figure 2b arise from the A mode, whereas those above it arise from the S mode. Figure 2b, 2c and 2d show that S mode exhibits relatively lower intensity, which is because of the finite slit width, as discussed above.

Having identified the origins of the *p*- and *s*-polarized components (arising from TR and SPP-induced 1D PlC modes, respectively), we thereby investigate the CPL generation due to the interference of these two components, as described in Figure 1a. Figures 2e-2h confirm the generation of both LCP (blue) and RCP (red). Figures 2e and 2f show the CPL dispersion relations with the excitation at the left and right sides of the terrace, respectively (see insets). In both figures, CPL handedness inversion is observed below the calculated dispersion line (solid black line). To assess the position dependent CPL and handedness inversion, an energy-scan $S_3$ (RCP-LCP signal) line profile (Figure 2g) is employed, under the same configuration as in Figure 2c, at emission angle $\theta = 20°$. Figure 2g clearly shows that the handedness inversion occurs along the terrace edges and at the resonance energy of A mode (orange dash line). The CPL handedness is determined by the phase difference between the *p*- and *s*-polarized components. In this system, the phase of the *p*-polarized component originating from TR can be considered nearly constant.[27] Consequently, the phase of the *s*-polarized component can be extracted from the CPL handedness. The observed handedness inversion can therefore be explained by the dipole-like field formed at the terrace edge flipping its phase; a $\pi$ phase shift of *s*-polarized light occurs due to the dipole edge or to the resonance. As a result, the CPL generated at the opposite side of the edges exhibits opposite handedness as well as at the energies above and below the resonance. This is directly related to the dispersion curve of Figure 2e and 2f, where the opposite CPL handedness exhibits the symmetry with respect to the central axis of the terrace, further confirming the phase variation of the *s*-polarized component. Moreover, when the energy increases to the resonance of S mode, no CPL intensity is visible at the center of terraces in Figure 2d, indicating that S mode dose not contribute to CPL generation. However, CPL intensity is still visible at the terrace edges, which is attributed to the residual intensity of the A mode. Therefore, in Figures 2g, the handedness remains unchanged when the energy exceeds to the resonance energy of the S mode. At the higher energies ($> 3$ eV), the same types of phase flips of the higher order modes are also visible.

Figure 2h presents emission angle-resolved $S_3$ line profiles of a single terrace, showing how the CPL emission on the terrace varies with energy and emission angle. The solid line attached with pink dots is the calculated dispersion line, and the orange dots denote the resonance peaks of the A mode at each emission angle, thereby indicating its dispersion. Accordingly, Figure 2h reveals that a $\pi$ phase difference exists between the *s*-polarized component in the energy range above and below the orange-dot dispersion line of A mode. These results demonstrate that the phase of *s*-polarized component can be extracted by leveraging *p*-polarized light (TR) as a reference field. Moreover, the phase can be dynamically controlled by adjusting the electron beam position, energy, and detection angle, thereby enabling flexible modulation of CPL generation, i.e., the SAM of light.

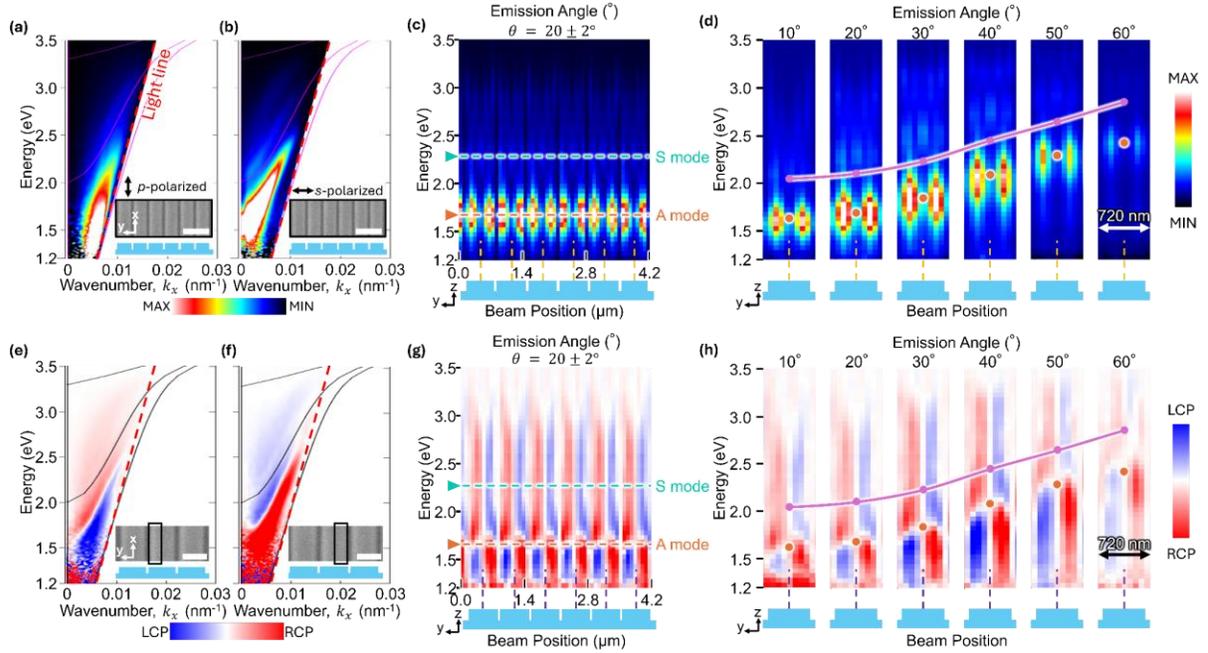

**Figure 2.** Results of the large terrace 1D PlC. Dispersion plots for the (a) *p*-polarized and (b) *s*-polarized light, with the backscattered electron images shown in the insets. The calculated SPP dispersion curves are represented by pink solid lines, and the light line is shown as a red dashed line. (c) *S*-polarized energy-scan line profiles across the sample periodicity (along the *y* direction) acquired at a detection angle of $\theta = 20° \pm 2°$. (d) *S*-polarized energy-scan line profiles of a single terrace at different angles. $S_3$ (difference of RCP and LCP intensities) dispersion plots for the (e) left-edge- and (f) right-edge-excitation of the terrace, with backscattered electron images in the insets. The calculated SPP dispersion curves are indicated by black line and the light line by red dashed line. (g) Energy-scan $S_3$ line profiles collected at $\theta = 20° \pm 2°$. (h) Energy-scan $S_3$ line profiles of a single terrace at different angles. The scale bars in panels a and b represent 1 μm, while those in panels e and f represent 500 nm. In panels c, d, g, and h, the vertical dashed line marks the position of the terrace center. In panels d and h, the calculated dispersion curve in the angular space is depicted by pink dots with connected lines, and the orange dots represent the resonance peak of A mode.

**CPL generation in the 1D PlC with small terrace (120 nm)**

We now extend the above discussed mechanism of CPL generation from the large terrace sample (terrace width of 420 nm) to the small terrace sample (terrace width of 120 nm) where we have to consider the emission from the groove as well. Figure 3 shows CL measurement results of the small-terrace sample. Figures 3a and 3b show the *p*- and *s*-polarized dispersion plot, respectively. The CL intensity is integrated over the entire scanning area. For the *p*-polarization in Figure 3a, the detected emission aligned along the light line indicates that the *p*-polarized emission is dominated by TR, similar to the large terrace sample. For the *s*-polarization in Figure 3b, the emissions follow the calculated dispersion curves (shown by solid pink curves), revealing the dispersive features of the 1D PlC modes. A splitting of the bright pattern along the dispersion curve is also observed, similar to that of the large-terrace sample.

Figure 3c shows an energy-scan line profile of *s*-polarized emission with the emission angle of 20°. For the small-terrace sample, the S mode (with charges at the center of terrace and groove) appears at lower energy, while the A mode (with charges at the terrace edges) occurs at higher energy. While this energy ordering is opposite to that observed in the large-terrace sample (Figure 2c)[36], the two modes still align well with the calculated dispersion line, as shown in Figure 3d, indicating their dispersive features are caused by propagating SPPs.

Next, we investigate CPL generation from this small terrace structure. Figures 3e and 3f show the CPL dispersion plots for excitation at the left and right sides of the terrace, respectively. In each dispersion plot, only a single polarization state is observed with the opposite handedness for the left- and right-side excitations. The $S_3$ energy-scan line profile at an emission angle of 20° (Figure 3g) shows only LCP (blue) and RCP (red) emissions at left and right terrace edges, respectively, indicating that the polarization states at these locations remains constant across the measured energy range while retaining opposite handedness for the left and right terrace edges. Figure 3h presents emission angle-resolved $S_3$ energy-scan line profiles for a single terrace, showing that LCP is generated at the left terrace edge, whereas RCP is generated at the right terrace edge, regardless of energy or emission angle. These observations indicate that the phase of *s*-polarized component is retained nearly constant over the emission angle and energy, only with position dependence. This CPL behavior cannot be explained solely by A mode as discussed above for the large terrace sample. In this structure, the *s*-polarization component is instead dominated by emission from the localized dipolar mode at terrace edges, hereafter referred to as local mode. Its emission exhibits no angular dependence around 1.5 eV with charge accumulation at the edges. (see Figure S1 in Supporting Information (SI)) Because of the broad spectral bandwidth of the localized dipole, the radiation phase varies weakly with energy and emission angle, and consequently the CPL produced by interference with TR has a handedness that is independent of both energy and emission angle. This interpretation is consistent with Figure 3h, where the polarization at the terrace edges remains unchanged across the measured energy and emission angle ranges. Thus, under this condition with the electron beam on the terrace, the CPL handedness is sensitive only to the excitation position along *y* direction, but not to the angle or energy.

At the energies above the resonance peak of A mode (above the orange dots in Figure 3h), a pronounced reduction in CPL intensity is observed. This behavior can be explained by considering the two interference pathways that contribute to the CPL signal: Below the A mode resonance, the CPL arising from TR–A mode interference and TR–local mode interference has the same handedness, leading to constructive sum and relatively strong emission. However, above the resonance energy, the CPL generated by TR–A mode interference reverses its handedness (due to the phase flip of A mode) and becomes opposite to that generated by TR–local mode interference. As a result, CPL from these two sources cancel each other, leading to a sudden reduction in the overall CPL intensity at the A mode resonance energy, as shown in Figure 3h.

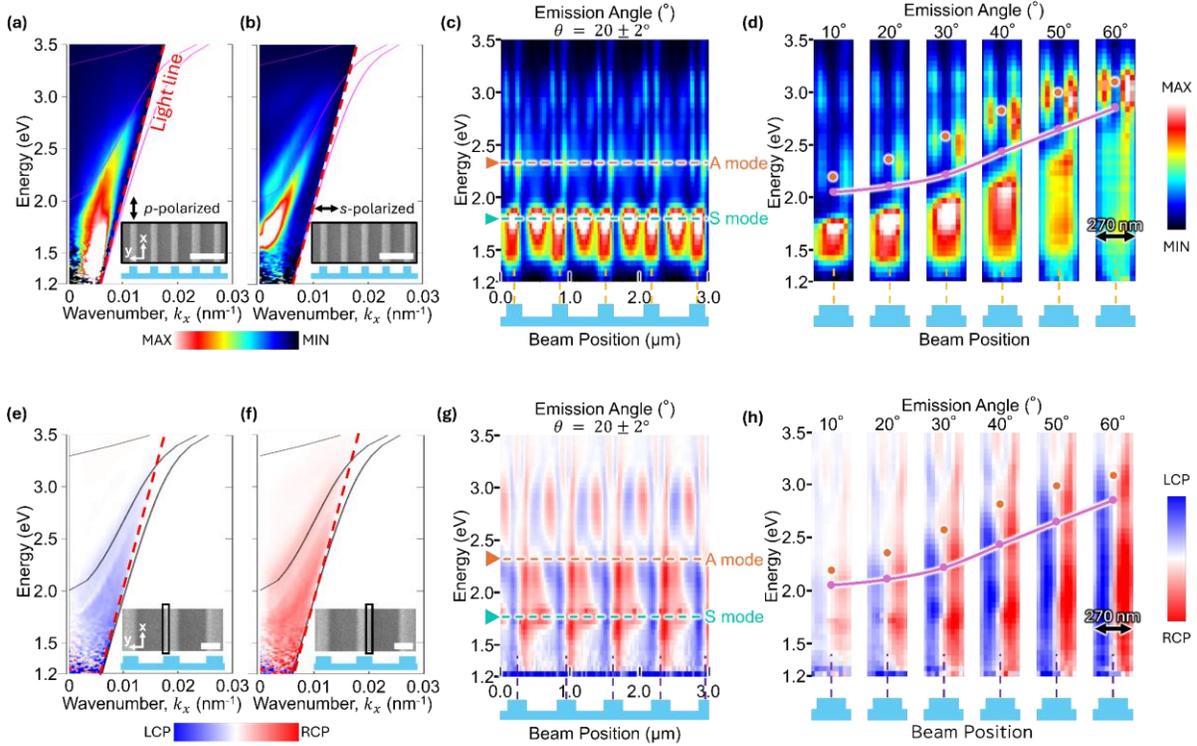

**Figure 3.** Results of the small terrace 1D PlC. Dispersion plots for the (a) *p*-polarized and (b) *s*-polarized light, with the backscattered electron images in the insets. The calculated SPP dispersion curves are represented by pink solid lines, and the light line is shown as a red dashed line. (c) *S*-polarized energy-scan line profiles across the sample periodicity (along the *y* direction) acquired at a detection angle of $\theta = 20° \pm 2°$. (d) *S*-polarized energy-scan line profiles of a single terrace at different angles. $S_3$ dispersion plots for the (e) left-edge- and (f) right-edge-excitation of the terrace, with backscattered electron images in the insets. The calculated SPP dispersion curves are indicated by black line and the light line by red dashed line. (g) Energy-scan $S_3$ line profiles

collected at $\theta = 20° \pm 2°$. (h) Energy-scan $S_3$ line profiles of a single terrace at different angles. The scale bars in panels a and b represent 1 μm, while those in panels e and f represent 500 nm. In panels c, d, g, and h, the vertical dashed line marks the position of the terrace center. In panels d and h, the calculated dispersion curve in the angular space is depicted by pink dots with connected lines, and the orange dots represent the resonance peak of A mode.

As shown in the line profile of Figure 3c, the significant portion of the CL emission in the small terrace sample originates also from the groove positions. At the resonance energy of the S mode, emission is concentrated at the terrace and groove center, consistent with the charge distribution of the mode. At the A mode resonance, the groove emission splits toward the groove edges and exhibits a charge distribution similar to that of the A mode on the terrace. Therefore, emission from the groove is also important for the small terrace sample. As shown in the emission angle-resolved energy-scan line profile of *s*-polarized light (Figure 4a), the groove emissions from the small terrace sample also follow the calculated dispersion line (solid pink curve), indicating a pronounced dispersive feature caused by propagating SPP. We therefore also perform the CPL analysis on emissions from the groove, as shown by the emission angle-resolved $S_3$ energy-scan line profiles extracted from the single groove region (Figure 4b). Around the A mode resonance (orange points), the groove CPL emission is like *s*-polarized dipolar type radiation with spatially split handedness at the sides. Also, across the A mode resonance energy, the CPL handedness flips indicating the $\pi$ phase difference (Figure 4b). This $\pi$-phase flip across the A mode resonance is clearly observable unlike the terrace position (Figure 3h) because of the absence of the local mode. The observed CPL emission at the S mode resonance energy (green points) is attributed to slight sample misalignment; the terrace/groove lines are not perfectly aligned along the *x* axis. Indeed, the CPL handedness at this position flips by slightly changing the sample orientation angle around the *x* axis (see Figure S2 in SI).

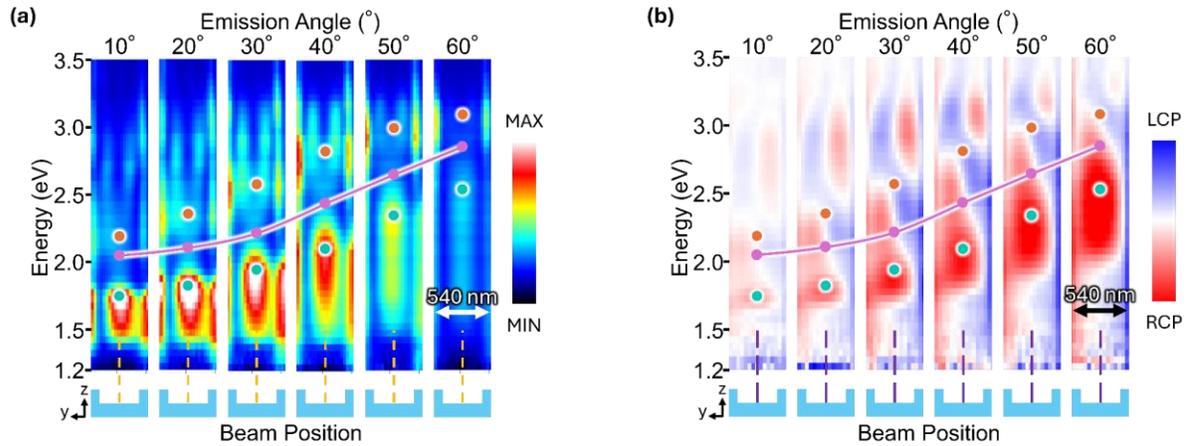

**Figure 4.** Results of the single groove of small terrace sample. (a) *S*-polarized energy-scan line profiles at different emission angles $\theta$. (b) $S_3$ energy-scan line profiles at different angles. The vertical dashed line marks the center position of the groove. The calculated dispersion curve in the angular space is depicted by pink dots with connected lines, and the orange and green dots represent the resonance peak of A mode and S mode, respectively.

**Emission modulation at structure boundary (small terrace)**

While we have discussed the interference of the 1D PlC mode and TR, we can also introduce another coherent emission source at the sample edge, where the structure ends. For the sample orientation of this measurement, the SPP component propagating along the groove or terrace (*x* direction) is not modulated by the sample periodicity, therefore cannot directly couple to free space, making no contribution to the emission. However, at the boundary where periodic-structure region meets planar surface (Figure 5a), the SPPs propagating along the *x* direction can be scattered as photons. These scattered photons can coherently interfere with the CPL generated within the periodic-structure region, producing interference fringes in the spatial mapping.[37–39] The interference fringes spacing, $R$, is expressed as:[27,35]

$$R = \frac{2\pi}{k_{SPP} \pm k \sin\theta} \quad (1)$$

where $k_{SPP}$ represents the SPP wave vector, $k$ represents the scattered light wave vector and $\theta$ is the emission angle in polar direction (the bottom left of Figure 1c). The sign is determined by the edge geometry: " + " sign corresponds to the bottom (*x* negative) edge region, whereas " − " sign corresponds to the top (*x* positive) edge region, reflecting the SPP propagating in the negative and positive *x*-directions at the two edges, respectively. In this expression, the denominator represents the path-length difference between the two radiation components. We here focus on the bottom edge of the small terrace sample as shown in Figure 5a, where the scattering at the structure boundary is most significant because of the larger structure "contrast" (the flat area has the height of the terrace) and of the CL detection configuration. (see Figure S3 in SI)

Figure 5 shows the CL measurement and simulation results in the bottom edge area. To investigate the SPP along the terrace, we analyze the profiles along the left and right edges of the terrace, as indicated by A and B in Figure 5a. Figures 5b and 5c present angle-scan line profiles along the *x* direction in the regions A and B at an energy of 2.0 eV, where the intensity is integrated horizontally in each region (along the *y* direction). The purple arrows indicate the position of the structure boundary. In the line profiles (Figure 5b, c, f and g), we removed the local modulations originating from sample imperfection by normalizing the entire signal (corresponding raw data and normalization details are shown in Figure S4a–d in SI). Owing to the symmetry of the sample, the angle-scan line profiles of regions A and B (Figure 5b and 5c) exhibit interference patterns with identical fringe spacing but opposite CPL polarizations. In both profiles, the fringe spacing strongly depends on the emission angle. As indicated by Eq. (1), at low emission angles (perpendicular to the surface), due to the small path-length difference between the two radiation sources, a large fringe spacing is observed. In contrast, at high emission angles (parallel to the surface), the path-length difference increases, resulting in a smaller fringe spacing as observed. To perform a more direct comparison with the experimental results, we simulated the angle-scan line profile based on Eq. (1), as shown in Figure 5d. The simulation well reproduces the experimental results of Figure 5b and 5c, supporting the validity of this interference model. Details of the simulation are provided in the Methods section. We also notice discontinuity of intensity around $\theta = 20°$ in both angle-scan line profiles, which corresponds to the bandgap between A and S modes where the CPL intensity suddenly decrease. To compare the profiles of the edges A and B, as shown in Figure 5e, $S_3$ line plots at $\theta = 20°$ are extracted from Figures 5b and 5c. Due to the structural symmetry of the sample, the two curves (regions A and B) are nearly mirror-symmetric about the $S_3 = 0$ line, indicating equal emission intensity but opposite CPL handedness, consistent with Figure 5b and 5c. Pronounced oscillations with alternating peaks and dips are observed in two profiles, reflecting the spatial modulation of the CPL signal by constructive and destructive interference.

According to Eq. (1), fringe spacing $R$ is also dependent on the energy, or $k_{SPP}$ and $k$. Figures 5f and 5g present normalized energy-scan line profiles for region A and B at the emission angle of 20°. Both profiles show large fringe spacing at low photon energies and smaller fringe spacing at higher energies, indicating a strong energy dependence. For comparison, the profile is simulated based on Eq. (1), as shown in Figure 5h. The simulated pattern well reproduces the experimental results of the decreasing fringe spacing with increasing photon energy. While not included in the simulation, the experimental profiles (Figure 5f, g) show low-intensities around $E = 2.0$ eV, which correspond to the bandgap between the A and S modes, as also discussed for the angle-scan line profiles (Figures 5b and 5c).

In contrast to the bottom edge of the structure, the fringe contrast in the top-edge region is much weaker (see Figures S5 in SI). This is because the SPP scatters "backwards" at this boundary as the structured areas are lower in height. Since the CL signal detection by the parabolic mirror is only at the *x* positive direction (Figure S3 in SI), the contribution of the edge scattering is small for the top-edge configuration. Interference patterns at both the top and bottom edges of the large-terrace sample are also presented in Section VI of SI, where the phase flip of the A mode is also clearly observed by CPL in the interference fringe line.

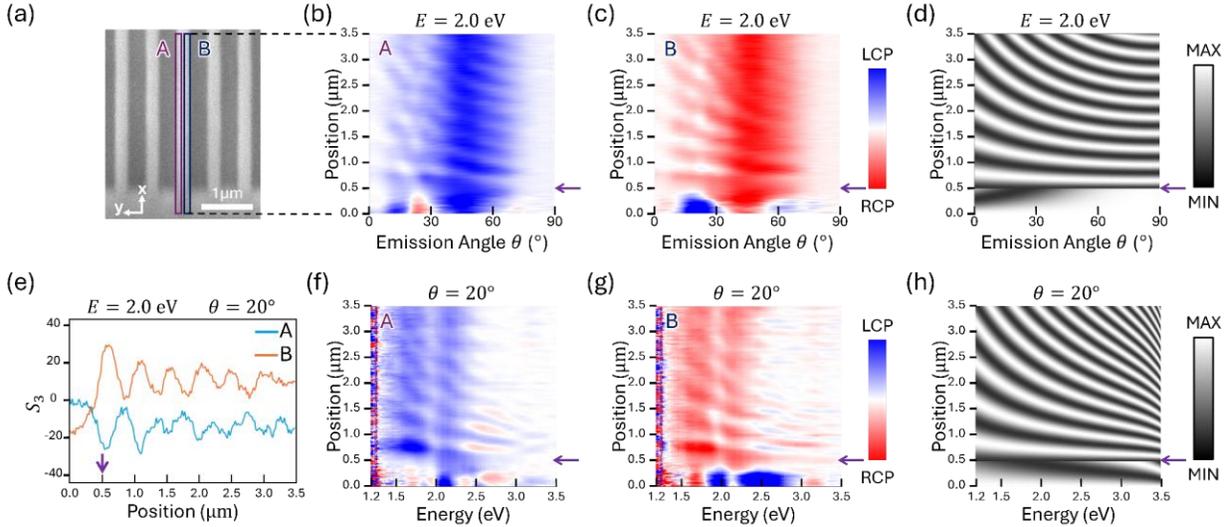

**Figure 5.** Results of the bottom edge of small terrace 1D PlC along *x* direction. (a) Secondary electron images for the bottom edge region. Region A and B are illustrated by the dark pink and dark blue boxes, respectively. Angle-scan $S_3$ line profiles of (b) region A, (c) region B and (d) simulation based on Eq. 1 collected at $E = 2.0 \pm 0.05$ eV. (e) $S_3$ line plots of regions A and B

collected at $E = 2.0 \pm 0.05$ eV and $\theta = 20° \pm 2°$. Energy-scan $S_3$ line profiles of (f) region A, (g) region B and (h) simulation based on Eq. 1 collected at $E = 2.0 \pm 0.05$ eV. In panels b, c, f, and g, the intensity of selected region is horizontally (along *y* direction) integrated. The purple arrows indicate the position of edges.

## CONCLUSION

In conclusion, we have demonstrated control of CPL in 1D PlCs with both large (420 nm) and small (120 nm) terrace width by using 4D STEM-CL measurement. We note that the CPL signal in CL corresponds to the field intensity when CPL is illuminated on the sample, thus presenting the asymmetric field distribution under CPL illumination in the purely light optical setup.

For the large-terrace sample, TR provides a *p*-polarized component that serves as a broadband reference field with a constant phase, while the *s*-polarized component originates from A mode. The CPL arises from the interference between these two components. Owing to the charge distribution of A mode (located at the terrace edges), a $\pi$ phase shift of *s*-polarized light occurs due to the dipole edge or to its resonance. Thus, the CPL handedness in large-terrace sample can be manipulated by excitation position, energy and emission angle. For small-terrace sample, it provides two distinct emission channels that enable complementary strategies for CPL control. Emission from the groove region reproduces the mechanism observed in large-terrace samples, in which the CPL handedness exhibits dispersion and excitation-position dependences. However, by preserving the sample periodicity while varying the terrace width, the A mode dispersion is shifted in energy, enabling control of the phase-flip position of the *s*-polarized component within the energy range and thereby modulating the energy at which the CPL handedness inversion occurs. Moreover, emission collected on the terrace yields a *y*-position-resolved CPL response dominated by local mode. Finally, in the bottom-edge region, interference with SPP scattering enables the efficiency of CPL modulation to be tuned via the *x*-position. The coexistence of these mechanisms allows flexible control of the SAM of light (CPL) through the choice of excitation position, photon energy, and emission angle, providing an effective route for dynamic manipulation of CPL at nanoscale. And inversely, by CPL illumination on such a sample, one can selectively excite nanoscopic positions asymmetrically with the CPL handedness.

## Methods

**4D STEM-CL Measurement System.** A modified STEM (JEM-2100F, JEOL, Japan) with a Cs-corrector is employed for the CL measurement and operated with an accelerating voltage of 80 kV. The electron probe is about 1 nA with a 20 mrad illumination half angle, resulting in a 1 nm probe size, which is optimized to excite signal precisely without observable damage to the sample. As shown in Figure 1c, a parabolic mirror is set in the STEM column to collimate the CL radiation from the sample placed at the focal point and bring it to the detection system outside the vacuum of the STEM instrument. In the optical path, a quarter-wave plate (QWP) and a linear polarizer are placed. A slit mask is set to fix the azimuthal angle at $\varphi = 0°$ while being able to collect all angle information in the spectrometer as 4D data sets with the electron beam scan information.[25,34,35] The 4D dataset (angle, energy and 2D space) is analyzed by custom-made software.

**Sample Fabrication.** A 1D periodic structure with a period $P = 600$ nm was fabricated on a planar surface composed of a high etch resistance layer (ZEP520A) on an indium phosphide (InP) substrate by using electron beam lithography. The structured area has a size of 30 µm. As shown in Figure 1b, each terrace has a rectangular cross-section with a height $H = 100$ nm and widths $D = 420$ nm and $D = 120$ nm, named as the large-terrace and small-terrace samples, respectively. A 200 nm thick silver layer was deposited onto the patterned structure by thermal evaporation, thereby forming the 1D PlC. Silver was chosen as the metallic medium for supporting SPPs due to its low optical loss in the visible and near-infrared spectral ranges. Finally, the sample was affixed onto a TEM copper grid using carbon paste for insertion into the sample chamber in STEM.

**Simulation Model.** The phase difference between CPL generated within the periodic-structure region and SPP scattering at bottom edge of small-terrace 1D PlC can be extracted from Eq. (1) and is given by $\varphi = L(k_{SPP} + k \sin\theta)$, where $L$ is the SPP propagation length, $k_{SPP}$ is the SPP wave vector, $k$ is the scattered light wave vector and $\theta$ is the emission angle in polar direction. The intensity is then expressed as $I = 1 + \cos(\varphi + \pi)$. This is because the interference model contains only two components, such that the fringe pattern is determined solely by their phase difference. Therefore, the remaining term in the intensity expression can be approximated as a constant, which is taken here to be 1. Notably, SPP is backscattered at the bottom edge because the CL detection system collects emission only in the positive $x$ direction. Due to this backscattering process, an additional $\pi$ phase is introduced.[40] By taking this effect into account, the calculated emission angle-scan and energy-scan line profiles match well with the experimental results (see Figure 5d and 5g).

## ASSOCIATED CONTENT

**Supporting Information**

Data analysis of local mode, misalignment of small-terrace sample, results of the top edge of small terrace 1D PlC, raw data of the bottom edge of small terrace 1D PlC, illustrations of the interference at top and bottom edges area and results of the top and bottom edges of large terrace 1D PlC can be found in the Supporting Information. This material is available free of charge at http://pubs.acs.org.

## AUTHOR INFORMATION

**Corresponding Author**

Takumi Sannomiya − Department of Materials Science and Technology, Institute of Science Tokyo, Yokohama 2268503, Japan;

orcid.org/0000-0001-7079-2937; *Email: sannomiya@mct.isct.ac.jp

**Authors**

Yuxin Yang – Department of Materials Science and Technology, Institute of Science Tokyo, Yokohama 2268503, Japan

Izzah Machfuudzoh – Department of Materials Science and Technology, Institute of Science Tokyo, Yokohama 2268503, Japan

Qiwen Tan − Department of Materials Science and Technology, Institute of Science Tokyo, Yokohama 2268503, Japan; orcid.org/0009-0003-6655-820X

**Author Contributions**

The manuscript was prepared with contributions from all authors. Y.Y. and T.S. conceived the study, conducted the experiments, and performed the analysis. I.M. and Q.T. contributed to the analysis and manuscript revision. All authors approved the final version of the manuscript.

**Funding Sources**

This work is financially supported by JSPS Kakenhi (JP24H00400, JP26K17736), JST FOREST (JPMJFR213J), and JST CREST (JPMJCR25I3).**Notes**

The authors declare no financial conflicts of interest.

# ACKNOWLEDGMENT

We thank Dr. Naoki Yamamoto and Dr. Sotatsu Yanagimoto for valuable advices. The authors acknowledge the financial support from JSPS Kakenhi (JP24H00400, JP26K17736), JST FOREST (JPMJFR213J), and JST CREST (JPMJCR25I3).

# Supporting Information

# Modulation of Spin Angular Momentum of Emission in Symmetric 1D Plasmonic Crystals by Cathodoluminescence

*Yuxin Yang[1], Izzah Machfuudzoh[1], Qiwen Tan[1], Takumi Sannomiya[1,]\**

[1] Department of Materials Science and Engineering, School of Materials and Chemical Technology, Institute of Science Tokyo, 4259 Nagatsuta, Midori-ku, Yokohama 226-8503 Japan

**Corresponding author**

*Takumi Sannomiya (Email: sannomiya@mct.isct.ac.jp)

## I. Local mode

Figure S1a shows the dispersion plot of the emission measured at the terrace edge (see inset). At approximately 1.5 eV, an emission feature with no angular dependence is observed, indicating the presence of a local mode. Figure S1b presents the photon map at an energy of 1.5 eV and a detection angle of $\theta = 20°$. The intensity is confined to the terrace edges, indicating the presence of the corresponding charge distribution. Owing to the broad spectral bandwidth, the radiation phase of this local mode varies weakly with energy and emission angle. As a result, the handedness of the CPL generated by TR-local mode interference is independent of both energy and emission angle and remains only position dependence, as discussed in Figure 3e–3h.

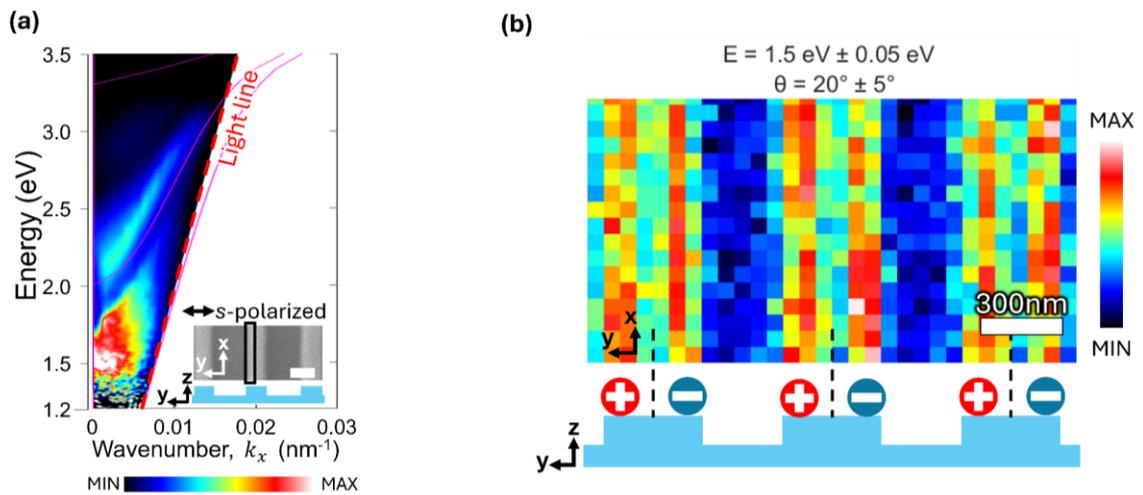

**Figure S1.** (a) Dispersion plot for the *s*-polarized light, presented along with the backscattered electron images in the insets. The calculated SPP dispersion curves are represented by pink solid lines, and the light line is shown as a red dashed line. The scale bar represents 500 nm. (b) Photon map acquired at energy of $E = 1.5 \text{ eV} \pm 0.05 \text{ eV}$ and detection angle of $\theta = 20° \pm 2°$. The illustration shows the charge distribution indicated by the photon map. The vertical dashed line marks the center position of the terrace.

## II. Misalignment of small-terrace sample

At the resonance energy of the S mode, Figure S2 shows LCP (blue) inside the groove, in contrast to Figure 3g and 4b of the main text, which show RCP (red). This is because the terrace/groove lines cannot be perfectly aligned along the $x$ axis, due to manual sample alignment. In the measurement shown in Figure S2, the sample is rotated in the opposite direction about $x$ axis relative to the measurements in Figures 3g and 4b, thereby introducing reversed asymmetry.

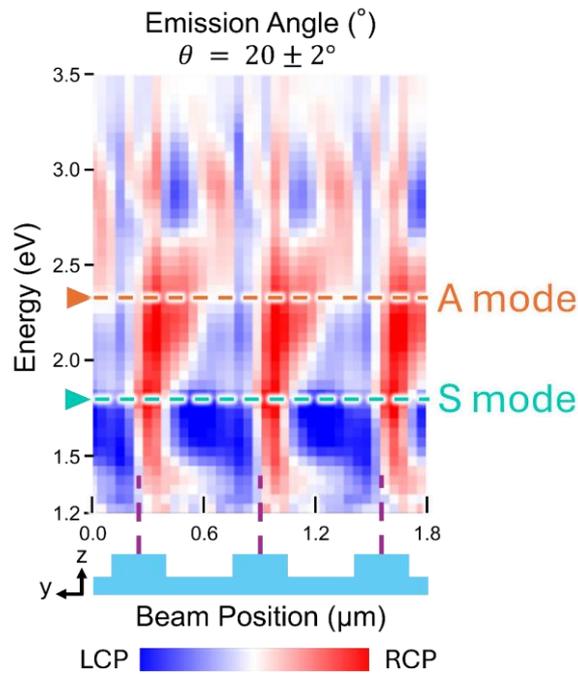

**Figure S2.** Energy-scan $S_3$ line profiles collected at $\theta = 20° \pm 2°$. It shows the result of sample with opposite rotation angle about the $x$ axis compared with Figure 3g and 4b in the main text. The vertical dashed line marks the center position of the terrace. The orange and green horizontal dashed line marks the resonance energy of A and S modes, respectively.

# III. Illustrations of the interference at top and bottom edges area

The difference in the interference observed at top and bottom edge regions are illustrated in Figure S3. SPPs excited by the electron beam propagate along the *x*-axis and are scattered at the groove edges. At the top edge, scattering from the groove toward the CL detector via the parabolic mirror, which is positive *x* direction, is blocked, thereby suppressing interference with the CPL generated within periodic-structure area, as shown in Figure S3a. In contrast, at the bottom edge, the backscattered photons of SPP can interfere with the CPL generated within periodic-structure area without being blocked (see Figure S3b), producing an interference-modulated CPL signal, as shown in Figure 5 of the main text.

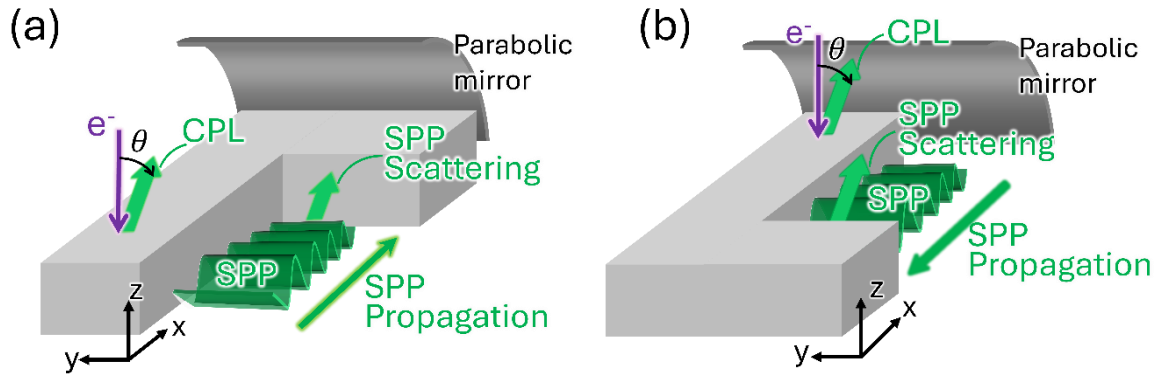

**Figure S3.** Illustrations of the interference, which occurs at the (a) top and (b) bottom edges area. $\theta$ denotes the polar emission angle. The SPP scattering at top edge (a) is blocked, resulting in an inefficient interference with the CPL produced within periodic-structure area.

# IV. Raw data of the bottom edge of small terrace 1D PlC

This section presents the raw data of the small terrace 1D PlC along *x* direction, presented in Figure 5 of the main text. Several fringes that are independent of both energy and emission angle can be observed in Figure S4, particularly in the high-angle regions of Figures S4a and S4b and near the 3.0 μm position in Figures S4c and S4d. These fringes are attributed to sample imperfections. To reduce the influence of these fringes, the angle-scan line profiles are first integrated over the emission angle, which averages out the interference contribution while retaining only the angle- and energy-independent fringes arising from sample imperfections. The resulting angle-integrated data are then used to normalize both the angle-scan and energy-scan line profiles shown in Figure 5 of the main text.

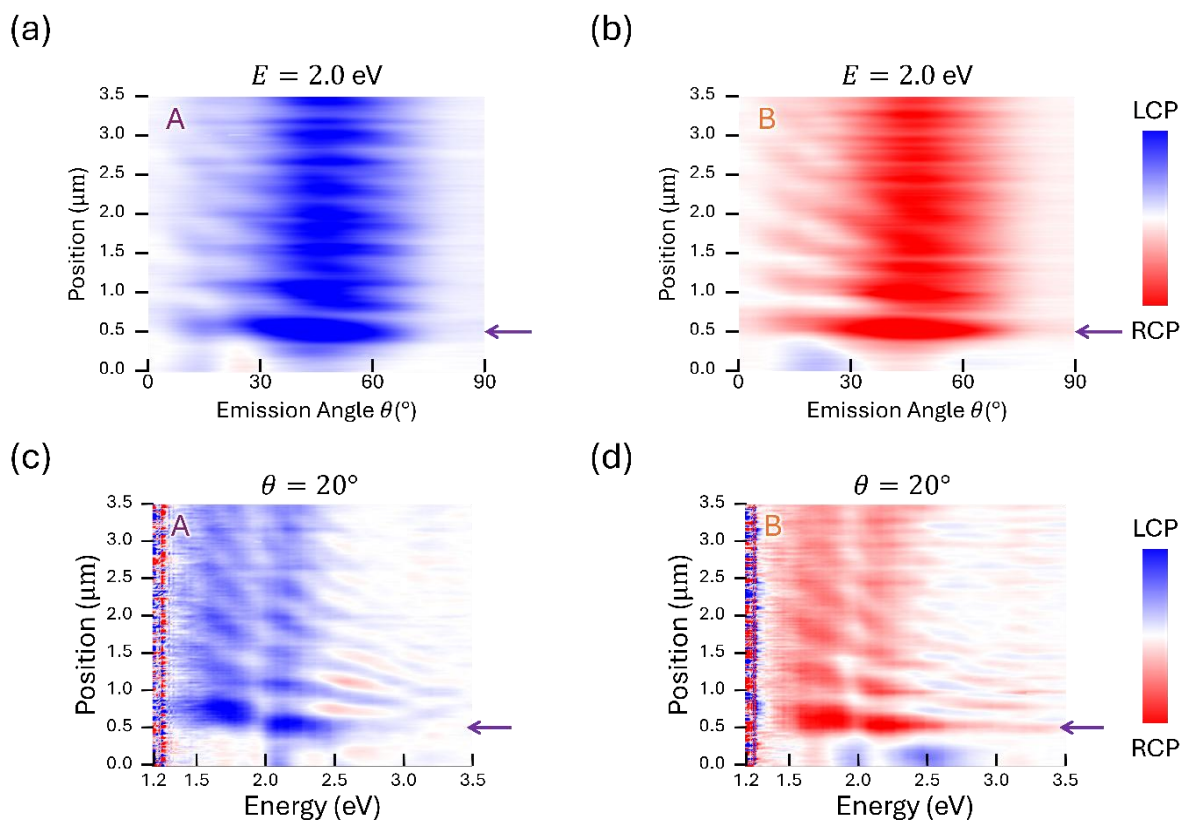

**Figure S4.** Raw data of the small terrace 1D PlC along *x* direction of Figure 5 in the main text. Angle-scan $S_3$ line profiles of region (a) A and (b) B collected at $E = 2.0 \pm 0.05$ eV. Energy-scan $S_3$ line profiles of region (c) A and (d) B collected at $E = 2.0 \pm 0.05$ eV. In panels a, b, c, and d, the intensity of selected region is integrated horizontally (along *y* direction). The purple arrows indicate the position of edges.

## V. Results of the top edge of small terrace 1D PlC

Figure S5 shows the result at the top edge of small terrace 1D PlC along *x* direction. Based on Eq. 1, the fringe pattern is expected to show both angular and energy dependence, regardless of whether the scattering originates from the top or bottom edge. However, as shown in Figure S5b and S5c, the fringe patterns in the top-edge region exhibit no angular dependence. We attribute the angle-independent fringes to reflected SPP waves by the top edge and to sample imperfections. Consistent with this interpretation, no pronounced fringe pattern is observed in Figure S5e and S5f. Figure S5d shows the $S_3$ line plots along the $x$ direction in region A and B after integrating the signal over the $y$ direction at $E = 2.0$ eV and $\theta = 20°$. The absence of clear peaks and dips indicates that constructive and destructive interference is not observed in this region, which is due to the SPP scattering originating from the groove is blocked by the top edge and thus cannot effectively interfere with the CPL within the periodic-structure region, as illustrated in Figure S3a.

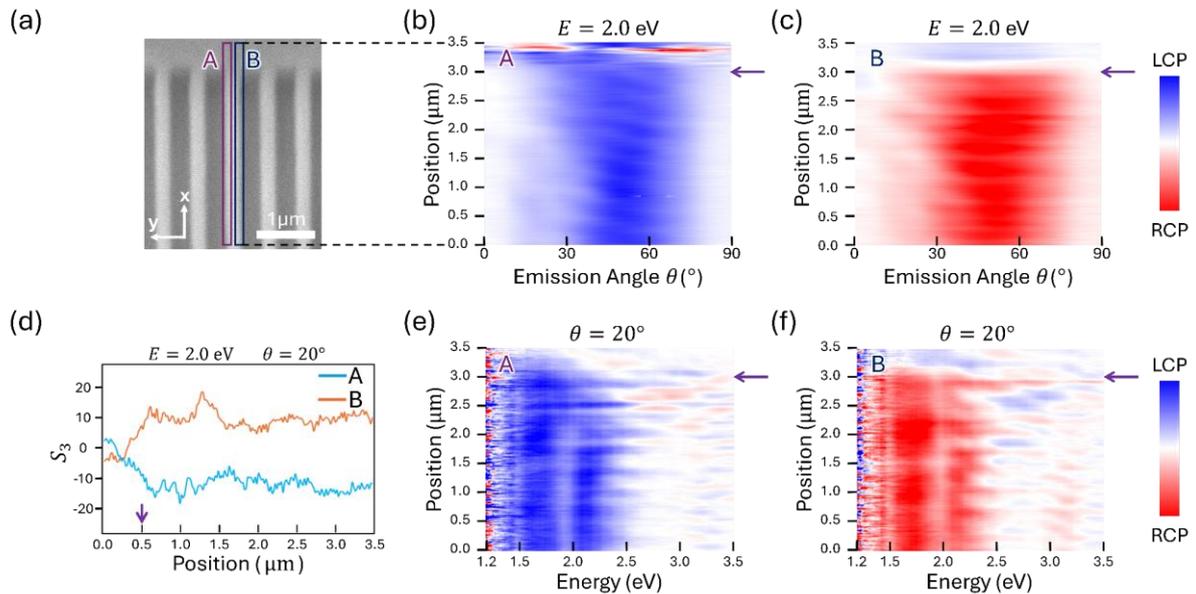

**Figure S5.** Results of the top edge of small terrace 1D PlC along $x$ direction. (a) Secondary electron images for the top edge region, with darker area corresponding to grooves. Region A and B are illustrated by the green and orange boxes, respectively. Angle-scan $S_3$ line profiles of (b) region A and (c) region B collected at $E = 2.0 \pm 0.05$ eV. (d) $S_3$ line plots of regions A and B collected at $E = 2.0 \pm 0.05$ eV and $\theta = 20° \pm 2°$. Angle-scan $S_3$ line profiles of (e) region A and (f) region B collected at $E = 2.0 \pm 0.05$ eV. In panels b, c, e, and f, the intensity of selected region is integrated horizontally (along $y$ direction). The purple arrows indicate the position of edges.

## VI. Results of the top and bottom edges of large terrace 1D PlC

Figures S6 and S7 show the results obtained at the top and bottom edges of large terrace 1D PlC along *x* direction. These results exhibit fringe patterns with little or almost no angular or energy dependence. This behavior is completely different from that predicted by the interference model described in Eq. 1. It may be attributed to the absence of emission inside the groove, as shown in Figure 2d. The fringes are likely produced by waves reflected from the edge or by sample imperfections.

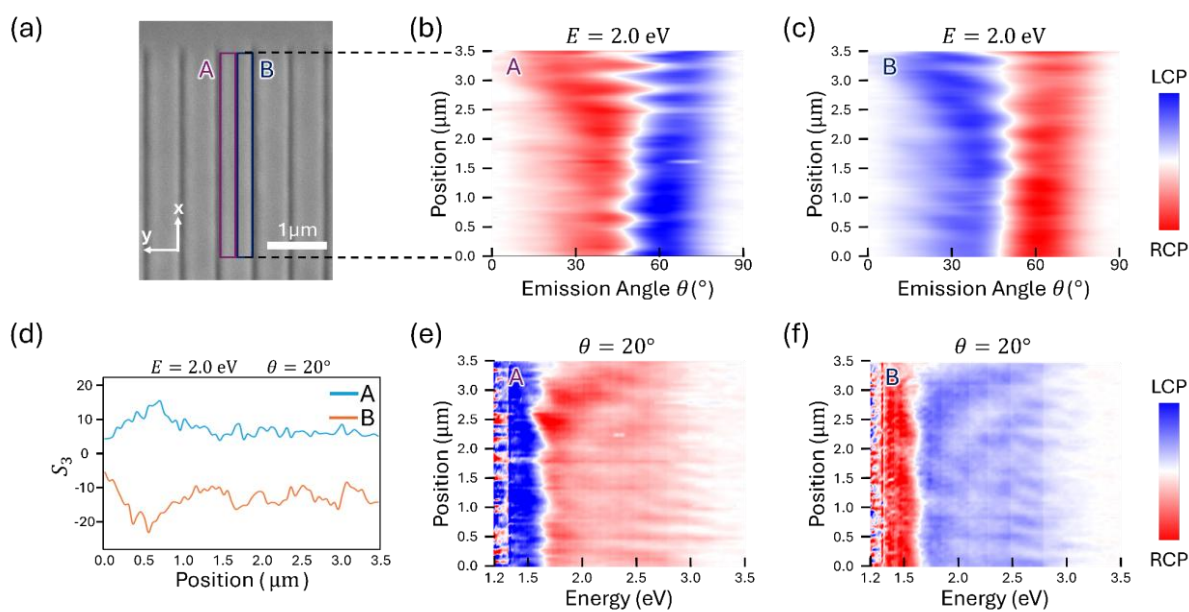

**Figure S6.** Results of the top edge of large terrace 1D PlC along *x* direction. (a) Secondary electron images for the top edge region. Region A and B are illustrated by the dark pink and dark blue boxes, respectively. Angle-scan $S_3$ line profiles of region (b) A and region (c) B collected at $E = 2.0 \pm 0.05$ eV. (d) $S_3$ line plots of regions A and B collected at $E = 2.0 \pm 0.05$ eV and $\theta = 20° \pm 2°$. Angle-scan $S_3$ line profiles of region (e) A and region (f) B collected at $E = 2.0 \pm 0.05$ eV. In panels b, c, e, and f, the intensity of selected region is integrated horizontally (along *y* direction).

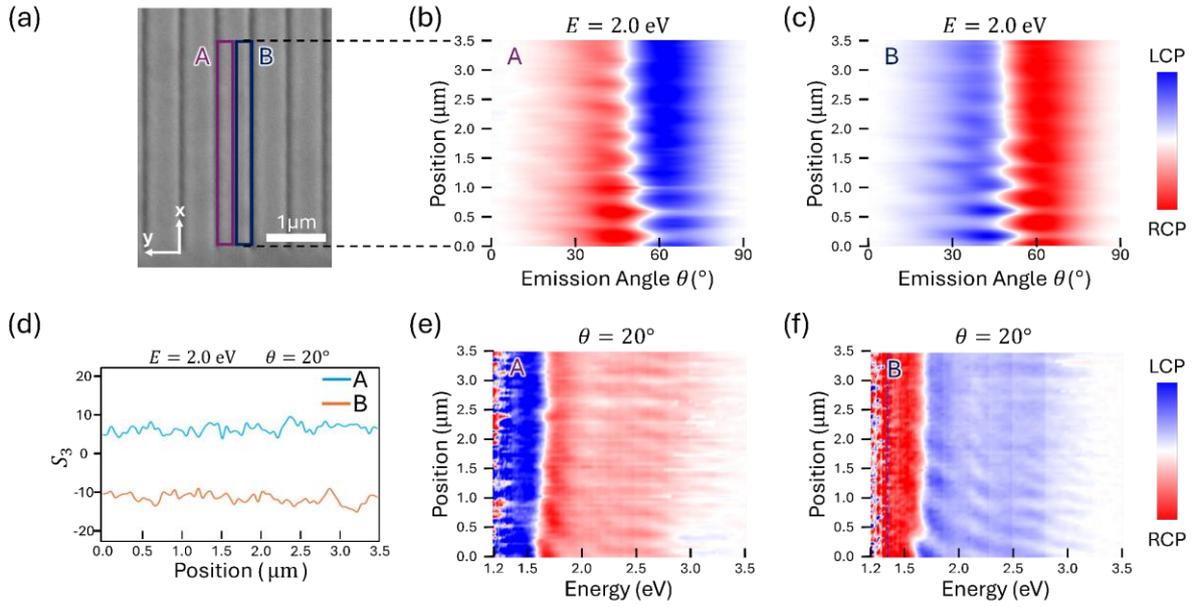

**Figure S7.** Results of the bottom edge of large terrace 1D PlC along *x* direction. (a) Secondary electron images for the bottom edge region. Region A and B are illustrated by the dark pink and dark blue boxes, respectively. Angle-scan $S_3$ line profiles of region (b) A and region (c) B collected at $E = 2.0 \pm 0.05$ eV. (d) $S_3$ line plots of regions A and B collected at $E = 2.0 \pm 0.05$ eV and $\theta = 20° \pm 2°$. Angle-scan $S_3$ line profiles of (e) region A and (f) region B collected at $E = 2.0 \pm 0.05$ eV. In panels b, c, e, and f, the intensity of selected region is integrated horizontally (along *y* direction).